\documentclass[conference]{IEEEtran}

\usepackage{cite,graphicx}
\usepackage{amsmath,amssymb,amsfonts}

\begin{document}

\title{Optimal Load Balancing in Millimeter Wave Cellular Heterogeneous Networks}

\author{\IEEEauthorblockN{Simin Xu, Nan Yang, and Shihao Yan}
\IEEEauthorblockA{Research School of Engineering, The Australian National University, Canberra, ACT 2601, Australia}Email: \{simin.xu, nan.yang, shihao.yan\}@anu.edu.au}

\markboth{Submitted to IEEE ICC 2018}{Xu \MakeLowercase{\textit{et
al.}}: Optimal Load Balancing in Millimeter Wave Cellular Heterogeneous Networks} \maketitle

\begin{abstract}
In this paper, we propose a novel and effective approach to optimizing the load balancing in a millimeter wave (mmWave) cellular heterogeneous network (HetNet) with a macro-tier and a micro-tier. The unique characteristics of mmWave transmission are incorporated into the network by adopting the Poisson point process (PPP) for base station (BS) location, the line-of-sight (LoS) ball model for mmWave links, the sectored antenna model for key antenna array characteristics, and Nakagami-$m$ fading for wireless channels. To reduce the load of macro-tier BSs, we consider a bias factor $A_{s}$ in the network for offloading user equipments (UEs) to micro-tier BSs. For this network, we first analyze the loads of macro- and micro-tier BSs. Then we derive a new expression for the rate coverage probability of the network, based on which the optimal $A_{s}$ maximizing the rate coverage probability is found. Through numerical results, we demonstrate the correctness of our analysis and the validity of the optimal $A_{s}$. Importantly, the optimal $A_{s}$ can bring a profound improvement in the rate coverage probability relative to a fixed $A_{s}$. Furthermore, we evaluate the impact of various network parameters, e.g., the densities and the beamwidths of BSs, on the rate coverage probability and the optimal $A_{s}$, offering valuable guidelines into practical mmWave HetNet design.
\end{abstract}

\IEEEpeerreviewmaketitle

\section{Introduction}\label{sec:intro}

Looking 3--5 years ahead, the fifth generation (5G) cellular networks will provide ultra-high quality of wireless service, such as up to gigabits per second throughput, towards a large number of mobile users. To bring such networks into reality, millimeter wave (mmWave) and network densification have been acknowledged as two highly promising techniques and received substantial interests from academia and industry \cite{overview1,overview4}. On one hand, the use of mmWave exploits the large available spectrum in the mmWave band. On the other hand, the use of network densification, i.e., deploying more base stations (BSs) in the network, significantly reduces the number of user equipments (UEs) which compete for the resources at each BS. Due to such potentials, the joint exploitation of mmWave and network densification has recently emerged as an active and quickly advancing research direction, the outcomes of which will help to alleviate the spectrum shortage problem facing global cellular providers and meet the rapidly growing data rate requirement.

To fully reap the benefits of mmWave communications, antenna arrays are employed at BSs for performing directional beamforming~\cite{hetnet9,hetnet7,blockModel2}. Thanks to the small wavelength of the mmWave band, it is possible to pack multiple antenna elements into the limited space at mmWave BSs, creating very narrow beams to provide very high gain. Against this background, numerous research efforts have been devoted to investigating mmWave wireless networks. For example,~\cite{overview12,channelmodel2} evaluated the channel characteristics of mmWave networks,~\cite{hetnet8,cover6} analyzed the achievable performance of mmWave networks, and~\cite{beamforming21,beamforming22} addressed the beam alignment problem in mmWave networks.

A highly effective mechanism to densify wireless cellular networks is to adopt the heterogeneous network (HetNet) architecture. The rationale behind this architecture is to deploy both macro BSs and low-power BSs, which differ not only in transmit power but in density, antenna array size, and height, to serve UEs. As pointed out by the 3rd generation partnership project (3GPP), a major issue in the HetNet is that macro BSs are often heavily loaded, while low-power BSs are always lightly loaded~\cite{load1}. This load disparity inevitably leads to suboptimal resource allocation across the network; therefore, a large number of UEs may be associated with one macro BS but experience poor date rate. To increase the load of low-power BSs and strike a load balance between macro BSs and low-power BSs, an association bias factor for low-power BSs needs to be added to increase the possibility that UEs are associated with low-power BSs~\cite{load1,load2}. This method, referred to as the cell range extension (CRE)~\cite{load1,load2}, offloads more UEs to low-power BSs, artificially expands the association areas of low-power BSs, and enables the network to better allocate its resource among UEs. Motivated by such potential, some studies (e.g.,~\cite{hetnet2,hetnet12}) evaluated the benefits of CRE for the performance of the sub-6 GHz cellular HetNet.

Our focus of this paper is to tackle a pressing challenge in the CRE optimization for mmWave cellular HetNets, i.e., determining the optimal load balancing. To the best knowledge of the authors, this challenge has not been addressed in the literature. We note that the previous studies have not examined the CRE in mmWave networks, e.g.,\cite{overview12,channelmodel2,hetnet8,cover6,beamforming21,beamforming22}, or investigated the CRE in sub-6 GHz cellular HetNets only, e.g.,~\cite{hetnet2,hetnet12}. We note that the results in~\cite{hetnet2,hetnet12} cannot be easily extended to mmWave networks, due to fundamental differences between mmWave HetNets and conventional HetNets. First, highly directional beams are applied in mmWave cellular HetNets, which significantly changes the interference behavior across the network. Second, the blocking effect in mmWave cellular HetNets reduces the received signal strength by tens of dBs. Thus, this effect cannot be ignored in mmWave cellular HetNets. Due to these reasons, new studies need to be conducted to evaluate and optimize the impact of CRE on the performance of mmWave cellular HetNets.

In this paper, we determine the optimal load balancing in a two-tier generalized mmWave cellular HetNet where the macro BSs in the macro-tier and the low-power BSs in the micro-tier co-exist to serve UEs. The generality of our considered network lies in the use of the Poisson point process (PPP) to model the location of BSs, the use of the LoS ball model to characterize the probability of a communication link being LoS, the use of the sectored antenna model at BSs to capture key antenna array properties, and the use of Nakagami-$m$ fading to model the wireless channel. We assume that a bias factor is used in the network to offload UEs to low-power BSs. For this network, we first analyze the loads of macro BSs and low-power BSs. Based on such analysis, we derive a new expression for the rate coverage probability of the network, which allows us to find the optimal bias factor which achieves the maximum rate coverage probability. Using numerical results, we show that the optimal bias factor brings about a significant improvement in the rate coverage probability relative to a fixed bias factor, especially when the rate threshold is in the medium regime, demonstrating the validity of our analysis. We further comprehensively examine the impact of network parameters on the rate coverage probability and the optimal bias factor.

\section{System Model}\label{sec:system}

\begin{figure}[t]
    \begin{center}
        \includegraphics[width=0.8\columnwidth]{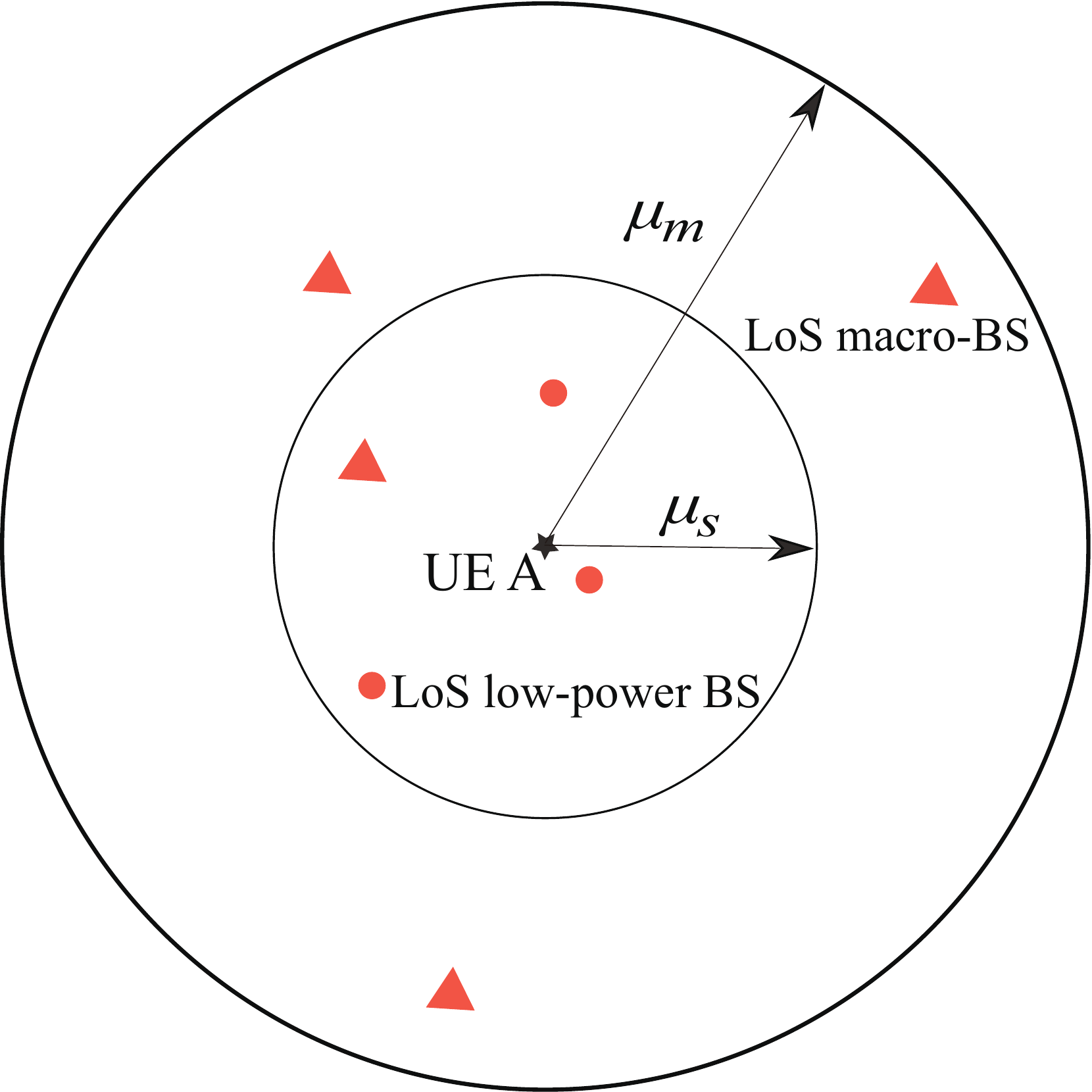}
        \caption{The illustration of the mmWave cellular HetNet under consideration. The typical UE, UE A, is located at the origin. All LoS macro BSs are located within $\mathcal{B}(0,\mu_m)$ and all LoS low-power BSs are located within $\mathcal{B}(0,\mu_s)$.}
        \label{fig:model}
    \end{center}
\end{figure}

We consider a mmWave cellular HetNet, as illustrated in Fig.~\ref{fig:model}, which consists of a macro-tier and a micro-tier. The macro-tier consists of macro BSs and the micro-tier consists of low-power BSs. In this work, we focus on the downlink of the HetNet where the BSs transmit to UEs. We model the macro BS location and the low-power BS location as two independent homogeneous PPPs, denoted by $\Phi_{all,m}$ with the intensity $\lambda_{all,m}$  on $\mathbb{R}^2$ and $\Phi_{all,s}$ with the intensity $\lambda_{all,s}$ on $\mathbb{R}^2$, respectively. We also model the UE location as another independent PPP, denoted by $\Phi_u$ with the intensity $\lambda_u$ on $\mathbb{R}^2$.

In this work, we randomly select one UE and refer to it as the typical UE (i.e., UE A in Fig.~\ref{fig:model}). Then we establish a polar coordinate system with the typical UE at the pole. Based on the Slivnyak theorem \cite{load2}, the properties observed at other UEs are the same as those observed at the typical UE. Therefore, we focus on the typical UE in this paper and analyze its rate coverage probability in the following sections. We clarify that the conclusions obtained for the typical UE can be extended to other UEs.

To establish a systematic performance study of the mmWave cellular HetNet, we need to incorporate the unique properties of mmWave communication into the network, as follows:

\subsubsection{Blockage Effect}

The impact of blockage caused by obstacles, e.g., buildings and trees, needs to be addressed in the network modeling. To this end, we adopt an accurate yet simple LoS ball model in the network, due to its high suitability for the system-level analysis of mmWave networks~\cite{cover6,hetnet9,hetnet7,blockModel1,blockModel2,cover3}. In the LoS ball model, the probability of a communication link being LoS is a function of the distance between the BS and the UE, $r$. In this work, we denote $P_{LoS,\xi}(r)$ as this probability, where $\xi=m$ for macro-tier and $\xi=s$ for micro-tier, and express $P_{LoS,\xi}(r)$ as
\begin{equation}\label{LoSBallModel}
P_{LoS,\xi}(r)=
\begin{cases}
\omega_\xi, & \mbox{if $0<r<\mu_\xi$}, \\
0, & \mbox{otherwise},
\end{cases}
\end{equation}
where $0\leq\omega_\xi\leq1$ is the value of the probability and $\mu_\xi>0$ is depicted in Fig.~\ref{fig:model}. According to \cite[Prop. (1.3.5)]{Book1}, the LoS macro BS location observed at the typical UE follows the PPP $\Phi_m$ with intensity $\lambda_{all,m}\omega_m$ within the circular area $\mathcal{B}(0,\mu_m)$. Moreover, the LoS low-power BS location observed at the typical UE follows the PPP $\Phi_s$ with intensity $\lambda_{all,s}\omega_s$ within the circular area $\mathcal{B}(0,\mu_s)$. These two PPPs are illustrated in Fig.~\ref{fig:model}. For the simplicity of presentation, we define $\lambda_{m}\triangleq\lambda_{all,m}\omega_m$ and $\lambda_{s}\triangleq\lambda_{all,s}\omega_s$.

\subsubsection{Directional Beamforming}

We assume that each UE is equipped with a single omnidirectional antenna. We also assume that each BS is equipped with an uniform planar square array with half-wavelength antenna spacing to perform directional beamforming. We further assume that the size of the array at the macro BS is larger than that at the low-power BS. In the considered mmWave cellular HetNet, each BS directs its beam to its associated UE for providing the best received signal quality. As adopted in \cite{hetnet9,hetnet7,blockModel2}, the actual antenna array gain pattern of BSs is approximated by a sectored antenna model, which captures the key antenna array characteristics including the beamwidth, the main lobe gain, and the side lobe gain. Mathematically, this pattern is expressed as
\begin{equation}\label{arrayGain}
G_{\xi}\left(\theta\right)=\begin{cases}
G_{\max,\xi}, & \mbox{if $|\theta|\leq\frac{\theta_\xi}{2}$,}\\
G_{\min,\xi}, & \mbox{otherwise},
\end{cases}
\end{equation}
where $G_{\max,\xi}$ and $G_{\min,\xi}$ are the main lobe gain and the side lobe gain, respectively, $\theta$ is the angle off the boresight direction, and $\theta_\xi$ is the beamwidth of the BS.

\subsubsection{Fading Channel Model}

In this work, we adopt independent Nakagami-$\mathcal{M}$ fading for each link to model the small scale fading $h$, where $\mathcal{M}$ is the Nakagami fading parameter. Accordingly, we define $\hbar\triangleq|h|^2$ as the small scale fading power gain. Therefore, $\hbar$ follows the Gamma distribution with the shape parameter $\mathcal{M}$ and the scale parameter $1/\mathcal{M}$. It follows that the cumulative distribution function (CDF) and the moment generation function (MGF) of $\hbar$ are given by
\begin{equation}\label{CDFofH}
F_{\hbar}\left(x\right)=1-e^{-\mathcal{M}x}\sum_{k=0}^{\mathcal{M}-1}\frac{(\mathcal{M}x)^k}{k!},
\end{equation}
and
\begin{equation}\label{MomentGeneration}
M_\hbar(s)=\left(1-\frac{s}{\mathcal{M}}\right)^{-\mathcal{M}},
\end{equation}
respectively, where $\mathcal{M}$ is assumed to be a positive integer.

We assume that all macro BSs have the same transmit power $P_m$ and all low-power BSs have the same transmit power $P_s$. Thus, the received power at the typical UE from a BS is given by $P=P_{\xi}G_{\xi}\left(\theta\right)r^{-\alpha}\hbar$, where $r$ is the distance between the BS and the typical UE and $\alpha$ is the path loss exponent.

We now present the UE association in the considered mmWave cellular HetNet. In this work, we assume that one UE is associated with only one BS. Due to the extremely high outdoor penetration loss of mmWave propagation~\cite{empirical1,overview4}, it is unpractical to rely on a NLoS BS to provide high data rate service. As such, we assume that only LoS BSs are qualified as the servers to transmit to UEs. Here, the maximum biased received signal strength (RSS) UE association algorithm~\cite{hetnet2} is adopted. Based on this algorithm, the typical UE is associated with either the nearest LoS macro BS or the nearest LoS low-power BS. We denote $A_s$ as the bias factor of the low-power BS tier and let $A_s\geq1$ to offload UEs to low-power BSs. We also denote $r_{m}$ as the distance between a LoS macro BS and the typical UE and $r_{s}$ as the distance between a LoS low-power BS and the typical UE. Therefore, the typical UE chooses the nearest LoS macro BS as its serving BS if
\begin{align}\label{condition_macro_choose}
P_{m}r_{\min,m}^{-\alpha}G_{\max,m}>&~A_{s}P_{s}r_{\min,s}^{-\alpha}G_{\max,s}\notag\\
\Rightarrow{}r_{\min,s}>&~\rho{}r_{\min,m},
\end{align}
where $r_{\min,m}=\min\{r_{m}\}$ is the distance between the nearest LoS macro BS and the typical UE, $r_{\min,s}=\min\{r_{s}\}$ is the distance between the nearest LoS low-power BS and the typical UE, and we define $\rho\triangleq\left({P_{m}G_{\max,m}}/{A_{s}P_{s}G_{\max,s}}\right)^{-{1}/{\alpha}}$ for the sake of simplicity. If $r_{\min,s}\leq\rho{}r_{\min,m}$, the typical UE chooses the nearest LoS low-power BS as its serving BS. As per the UE association algorithm, one BS can be chosen by multiple UEs at the same time. To avoid inter-user interference, we adopt the time-division multiple access (TDMA) scheme due to its popularity in the existing mmWave standards such as IEEE 802.11ad \cite{standard80211ad}, IEEE 802.15.3c \cite{standard802153c}, and ECMA-387 \cite{standardECMA387}.

\section{Load and Rate Coverage Probability Analysis}\label{sec:analysis}

In this paper, we aim to achieve the optimal load balancing in mmWave cellular HetNets. This aim mandates the characterization of the relationship between the bias factor, $A_s$, and the rate coverage probability, $P_{c}$. This characterization is completed within two steps. First, we derive the load of a typical macro BS and the load of a typical low-power BS, where the load of a BS is defined as the mean number of the UEs associated with this BS. Second, we analyze the rate coverage probability at the typical UE. We next detail the two steps in the following subsections.

\subsection{Load Analysis}\label{sec:load_analysis}

According to the Slivnyak theorem \cite{load2}, the load of a macro BS is given by $\lambda_{u}P_{t_{m}}/\lambda_{all,m}$, where $P_{t_{m}}$ is the probability that a UE is associated with a macro BS. Similarly, the load of a low-power BS is given by $\lambda_{u}P_{t_{s}}/\lambda_{all,s}$, where $P_{t_{s}}$ is the probability that a UE is associated with a low-power BS. For the convenience of presentation, we define $L_{m}\triangleq\lambda_{u}P_{t_{m}}/\lambda_{all,m}$ and $L_{s}\triangleq\lambda_{u}P_{t_{s}}/\lambda_{all,s}$. In order to evaluate $L_{m}$ and $L_{s}$, we need to derive $P_{t_{m}}$ and $P_{t_{s}}$, as follows.

We define $B_{m}\triangleq\Pr(\Phi_m\neq\emptyset)$ and $B_s\triangleq\Pr(\Phi_s\neq \emptyset)$. Based on \cite[Eq. (2.15)]{Book2}, we find that
\begin{equation}
B_m=1-e^{-\lambda_{m}\pi\mu_{m}^2}\notag
\end{equation}
and
\begin{equation}
B_s=1-e^{-\lambda_{s}\pi\mu_{s}^2}.\notag
\end{equation}
We also find that $B_m$ and $B_s$ are not necessarily equal to 1, which implies that the typical UE is possible to find no LoS macro BSs or LoS low-power BSs to associate. Thus, we define five possible scenarios of the considered mmWave cellular HetNet as
\begin{itemize}
\item Scenario 1: $\Phi_m=\emptyset$ and $\Phi_s=\emptyset$.
\item Scenario 2: $\Phi_m\neq\emptyset$ and $\Phi_s=\emptyset$.
\item Scenario 3: $\Phi_m=\emptyset$ and $\Phi_s\neq\emptyset$.
\item Scenario 4: $\Phi_m\neq\emptyset$ and $\Phi_s\neq\emptyset$, while the typical UE is associated with the nearest LoS macro BS.
\item Scenario 5: $\Phi_m\neq\emptyset$ and $\Phi_s\neq\emptyset$, while the typical UE is associated with the nearest LoS low-power BS.
\end{itemize}

First, we note that in Scenario 2, the typical UE can only be associated with a macro BS, while in Scenario 3, the typical UE can only be associated with a low-power BS. Thus, we have
\begin{equation}\label{Ptm_Prob_definition}
P_{t_{m}}=\Pr\left(\mbox{Scenario 2}\right)+\Pr\left(\mbox{Scenario 4}\right),
\end{equation}
and
\begin{equation}\label{Pts_Prob_definition}
P_{t_{s}}=\Pr\left(\mbox{Scenario 3}\right)+\Pr\left(\mbox{Scenario 5}\right).
\end{equation}

In order to evaluate \eqref{Ptm_Prob_definition} and \eqref{Pts_Prob_definition}, we need to find the statistics of $r_{\min,s}$ and $r_{\min,m}$ (or equivalently, the statistics of $r_{\min,\xi}$ where $\xi\in\{m,s\}$ as used in \eqref{arrayGain}). Given $\Phi_{\xi}\neq\emptyset$, the CDF and the probability density function (PDF) of $r_{\min,\xi}$ are given by
\begin{equation}\label{Frmins}
F_{r_{\min,\xi}}(x)=
\begin{cases}
\frac{1-e^{-\lambda_{\xi}\pi{}x^2}}{B_\xi}, & \mbox{if $0\leq{x}\leq\mu_{\xi}$},\\
1, & \mbox{if $x>\mu_{\xi}$},
\end{cases}
\end{equation}
and
\begin{equation}\label{frmins}
f_{r_{\min,\xi}}(x)=
\begin{cases}
\frac{2\lambda_{\xi}\pi{}x{}e^{-\lambda_{\xi}\pi{}x^2}}{B_\xi}, & \mbox{if $0\leq{x}\leq\mu_{\xi}$}, \\
0, & \mbox{if $x>\mu_{\xi}$},
\end{cases}
\end{equation}
respectively.

Using these statistics, $P_{t_m}$ is obtained as
\begin{equation}\label{PtM}
P_{t_{m}}=B_m\left(1-B_s\right)+B_{m}B_{s}
\int_{0}^{\mu_s}f_{r_{\min,s}}(x)F_{r_{\min,m}}\left(\frac{x}{C}\right)dx.
\end{equation}
Similarly, $P_{t_s}$ is obtained as
\begin{equation}\label{PtS}
P_{t_{s}}=B_s(1-B_m)+B_{m}B_{s}
\int_{0}^{\mu_m}f_{r_{\min,m}}(x)F_{r_{\min,s}}\left(Cx\right)dx.
\end{equation}
Substituting \eqref{PtM} and \eqref{PtS} into $L_{m}=\lambda_{u}P_{t_{m}}/\lambda_{all,m}$ and $L_{s}=\lambda_{u}P_{t_{s}}/\lambda_{all,s}$, respectively, we obtain $L_m$ and $L_s$.

\subsection{Rate Coverage Probability Analysis}\label{sec:coverage_analysis}

In the considered mmWave cellular network with TDMA, the maximum rate at which the information can be transmitted by a macro BS or a low-power BS is
\begin{equation}\label{MaxRate}
R=\frac{S_{\xi}}{L_{\xi}}\log_{2}\left(1+\textrm{SINR}_{\xi}\right),
\end{equation}
where $S_{\xi}$ is the spectrum resource allocated to a macro BS or a low-power BS and $\textrm{SINR}_{\xi}$ is the SINR when the UE is associated with a macro BS or a low-power BS. We note that the typical UE is associated with neither a macro BS nor a low-power BS in Scenario 1. Thus, in the following analysis we only focus on Scenarios 2, 3, 4, and 5. The SINR coverage probability in each scenario is defined as the probability that the SINR is higher than an SINR threshold, $\tau$. Mathematically, the SINR coverage probabilities for Scenarios 2, 3, 4, and 5 are expressed as
\begin{equation}\label{cov_prob_def_P2}
P_2\left(\tau\right)\triangleq\Pr\left(\textrm{SINR}_{m}>\tau,\Phi_m\neq\emptyset,\Phi_s=\emptyset\right),
\end{equation}
\begin{equation}\label{cov_prob_def_P3}
P_3\left(\tau\right)\triangleq\Pr\left(\textrm{SINR}_{s}>\tau,\Phi_m=\emptyset,\Phi_s\neq\emptyset\right),
\end{equation}
\begin{equation}\label{cov_prob_def_P4}
P_4\left(\tau\right)\triangleq\Pr\left(\textrm{SINR}_{m}>\tau,\Phi_m\neq\emptyset,\Phi_s\neq\emptyset\right),
\end{equation}
and
\begin{equation}\label{cov_prob_def_P5}
P_5\left(\tau\right)\triangleq\Pr\left(\textrm{SINR}_{s}>\tau,\Phi_m\neq\emptyset,\Phi_s\neq\emptyset\right),
\end{equation}
respectively. Here, we clarify that the expressions for $\textrm{SINR}_{\xi}$ in Scenarios 2--5 are different from each other.

The rate coverage probability of the network is defined as the probability that the maximum rate of the network, $R$, is larger than the rate threshold, $\delta$, i.e., $P_{c}\triangleq\Pr\left(R>\delta\right)$. Based on \eqref{cov_prob_def_P2}, \eqref{cov_prob_def_P3}, \eqref{cov_prob_def_P4}, \eqref{cov_prob_def_P5}, and the relationship between $\delta$ and $\tau$ given by $\tau=2^{\delta{}L_{\xi}/S_{\xi}}-1$, we express $P_{c}$ as
\begin{align}\label{Rate}
P_{c}=&P_2\left(2^{\frac{\delta L_m}{S_m}}-1\right)+P_3\left(2^{\frac{\delta L_s}{S_s}}-1\right)\notag\\
&+P_4\left(2^{\frac{\delta L_m}{S_m}}-1\right)+P_5\left(2^{\frac{\delta L_s}{S_s}}-1\right).
\end{align}
In order to obtain $P_{c}$, we next derive $P_2\left(\tau\right)$, $P_3\left(\tau\right)$, $P_4\left(\tau\right)$, and $P_5\left(\tau\right)$.

\subsubsection{Analysis of $P_2\left(\tau\right)$}

In Scenario 2, the power of the received signal at the typical UE is given by $S=\kappa_{m}\hbar$,
where $\kappa_{m}=P_{m}r_{\min,m}^{-\alpha}G_{\max,m}$. Due to the tens of dB reduction in received power caused by blockage, the interference from NLoS BSs are ignored in this work~\cite{cover6}. The power of the interference at the typical UE is given by
\begin{equation}\label{Imast}
I=I_m^{\ast}\triangleq\sum_{(r_m,\theta)\in\Phi_m/(r_{\min,m},\theta)}
P_{m}r_{m}^{-\alpha}G_m(\theta)\hbar.
\end{equation}
Based on $S$ and $I$, we derive $P_{2}\left(\tau\right)$ as
\begin{align}\label{P2_derivation}
P_{2}\left(\tau\right)\stackrel{(a)}{=}&
\Pr\left(\Phi_{m}\neq\emptyset\right)\!\Pr\left(\Phi_{s}=\emptyset\right)\!
\Pr\left(\frac{S}{I_m^{\ast}+\sigma^{2}}\!>\!\tau|\Phi_{m}\neq\emptyset\right)\notag\\
=&\left(1-B_s\right)B_{m}\notag\\
&\times\mathbb{E}_{r_{\min,m},I_m^{\ast}}
\left[\Pr\left(\hbar>\frac{\tau(I_m^{\ast}+\sigma^{2})}{\kappa_{m}}|\Phi_{m}\neq\emptyset\right)\right]\notag\\
\stackrel{(b)}{=}&\left(1-B_s\right)B_{m}\mathbb{E}_{r_{\min,m},I_m^{\ast}}
\left[e^{-\frac{\psi_{m}(I_m^{\ast}+\sigma^{2})}{r_{\min,m}^{-\alpha}}}\right.\notag\\
&\times\left.\sum_{k=0}^{\mathcal{M}-1}\frac{\left(\psi_{m}\left(I_m^{\ast}+\sigma^{2}\right)\right)^k}{k!r_{\min,m}^{-\alpha{}k}}\right]\notag\\
=&\left(1-B_s\right)B_{m}\int_{0}^{\mu_m}\sum_{k=0}^{\mathcal{M}-1}
\mathbb{E}_{I_m^{\ast}}\left[e^{-\psi_{m}(I_m^{\ast}+\sigma^{2})x^{\alpha}}\right.\notag\\
&\left.\times\frac{\left(\psi_{m}(I_m^{\ast}+\sigma^{2}\right)x^{\alpha})^k}{k!}\right]f_{r_{\min,m}}(x)dx,
\end{align}
where $\sigma^{2}$ is the noise power at the typical UE and $\psi_{m}={\mathcal{M}\tau}/{P_{m}G_{\max,m}}$. In \eqref{P2_derivation}, the equality $(a)$ is obtained based on the independence between $\Phi_m$ and $\Phi_s$, and the equality $(b)$ is obtained based on the CDF of $\hbar$ given in \eqref{CDFofH}.

We now introduce the Laplace transform of $I_m^{\ast}+\sigma^{2}$ to obtain a simpler presentation of $P_2\left(\tau\right)$. Specifically, the Laplace transform of $I_m^{\ast}+\sigma^{2}$ is given by
\begin{equation}\label{Laplace}
\mathcal{L}_{I_m^{\ast}+\sigma^{2}}(a)=\mathbb{E}_{I_m^{\ast}}\left[e^{-a(I_m^{\ast}+\sigma^{2})}\right].
\end{equation}
Therefore, the $k$th derivative of \eqref{Laplace} is given by
\begin{equation}\label{LaplaceDerivation}
\mathcal{L}_{I_m^{\ast}+\sigma^{2}}^{(k)}(a)
=\mathbb{E}_{I_m^{\ast}}\left[e^{-a(I_m^{\ast}+\sigma^{2})}(-1)^k(I_m^{\ast}+\sigma^{2})^k\right].
\end{equation}
Substituting \eqref{LaplaceDerivation} in \eqref{P2_derivation}, we obtain
\begin{align}\label{P2L}
P_2\left(\tau\right)=&(1-B_s)B_m\int_{0}^{\mu_m}\sum_{k=0}^{\mathcal{M}-1}
\frac{\left(-\psi_{m}x^{\alpha}\right)^k}{k!}\notag\\
&\times\mathcal{L}_{I_m^{\ast}+\sigma^{2}}^{(k)}(\psi_{m}x^{\alpha})f_{r_{\min,m}}(x)dx.
\end{align}
As indicated by \eqref{P2L}, we need to find $\mathcal{L}_{I_m^{\ast}+\sigma^{2}}(a)$ to obtain $P_2\left(\tau\right)$. Based on \eqref{Laplace}, we obtain $\mathcal{L}_{I_m^{\ast}+\sigma^{2}}(a)$ as
\begin{align}\label{LImAST}
&\mathcal{L}_{I_m^{\ast}+\sigma^{2}}(a)
=e^{-a\sigma^{2}}\mathcal{L}_{I_m^{\ast}}(a)\notag\\
&\stackrel{(c)}{=}e^{-a\sigma^{2}}e^{-\lambda_{m}\int_{r_{\min,m}}^{\mu_m}r\int_{0}^{2\pi}
\left(1-\mathbb{E}_\hbar\left[e^{-\hbar{}a{}P_{m}G_{m}\left(\theta\right)r^{-\alpha}}\right]\right)
d\theta{}dr}\notag\\
&\stackrel{(d)}{=}e^{-a\sigma^{2}}e^{-\lambda_{m}\int_{r_{\min,m}}^{\mu_m}r\int_{0}^{2\pi}  \left(1-\left(1+\frac{a{}P_{m}G_{m}\left(\theta\right)r^{-\alpha}}{\mathcal{M}}\right)^{-\mathcal{M}}\right)
d\theta{}dr}\notag\\
&\stackrel{(e)}{=}e^{-a\sigma^{2}}e^{-\lambda_m\int_{r_{\min,m}}^{\mu_m}r\Omega_{m}dr},
\end{align}
where the equality $(c)$ is obtained based on \cite[Cor. (2.3.2)]{Book1}, the equality $(d)$ is obtained based on the MGF of $\hbar$ given in \eqref{MomentGeneration}, and  the equality $(e)$ is obtained based on the antenna gain function given in \eqref{arrayGain}. In \eqref{LImAST}, we define $\Omega_{m}$ as
\begin{align}\label{Dm}
\Omega_m=&\theta_m\left[1-\left(1+\frac{aP_{m}G_{\max,m}}{\mathcal{M}r^{\alpha}}\right)^{-\mathcal{M}}\right]\notag\\
&+\left(2\pi-\theta_m\right)\left[1-\left(1+\frac{aP_{m}G_{\min,m}}{\mathcal{M}r^{\alpha}}\right)^{-\mathcal{M}}\right].
\end{align}

\subsubsection{Analysis of $P_3\left(\tau\right)$}

In Scenario 3, the power of the received signal at the typical UE is given by $S=\kappa_{s}\hbar$ and power of the interference at the typical UE is given by
\begin{equation}\label{Isast}
I=I_s^{\ast}\triangleq\sum_{\left(r_s,\theta\right)\in\Phi_s/\left(r_{\min,s},\theta\right)}
P_{s}r_{s}^{-\alpha}G_s\left(\theta\right)\hbar.
\end{equation}
Following the derivation procedure of $P_2\left(\tau\right)$, we obtain $P_3\left(\tau\right)$ as
\begin{align}\label{P3L}
P_3\left(\tau\right)=&\left(1-B_m\right)B_s\int_{0}^{\mu_s}\sum_{k=0}^{\mathcal{M}-1}
\frac{\left(-\psi_s x^{\alpha}\right)^k}{k!}\notag\\
&\times\mathcal{L}_{I_s^{\ast}+\sigma^{2}}^{\left(k\right)}\left(\psi_{s}x^{\alpha}\right)
f_{r_{\min,s}}\left(x\right)dx,
\end{align}
where $\psi_s={\mathcal{M}\tau}/{P_{s}G_{\max,s}}$ and
\begin{equation}\label{LaplaceS}
\mathcal{L}_{I_s^{\ast}+\sigma^{2}}\left(a\right)=e^{-a\sigma^{2}}
e^{-\lambda_s\int_{r_{\min,s}}^{\mu_s}r\Omega_{s}dr}.
\end{equation}
In \eqref{LaplaceS}, we define $\Omega_s$ as
\begin{align}\label{Ds}
\Omega_s\triangleq&\theta_s\left[1-\left(1+\frac{aP_{s}G_{\max,s}}{\mathcal{M}r^{\alpha}}\right)^{-\mathcal{M}}\right]\notag\\
&+\left(2\pi-\theta_s\right)\left[1-\left(1+\frac{aP_{s}G_{\min,s}}{\mathcal{M}r^{\alpha}}\right)^{-\mathcal{M}}\right].
\end{align}

\subsubsection{Analysis of $P_4\left(\tau\right)$}

In Scenario 4, the power of the received signal at the typical UE is given by $S=\kappa_{m}\hbar$ and power of the interference at the typical UE is given by $I=I_m^{\ast}+I_s$, where $I_m^{\ast}$ is defined in \eqref{Imast} and $I_{s}$ is given by
\begin{equation}\label{Is}
I_s\triangleq\sum_{\left(r_s,\theta\right)\in\Phi_s}P_{s}r_{s}^{-\alpha}G_s\left(\theta\right) \hbar.
\end{equation}
Based on $S$ and $I$, we derive $P_4\left(\tau\right)$ as
\begin{align}
P_4\left(\tau\right)
=&\Pr\left(\textrm{SINR}_{m}>\tau,\Phi_m\neq\emptyset,\Phi_s\neq\emptyset,r_{\min,s}>\rho{}r_{\min,m}\right)\notag\\
=&\Pr\left(\textrm{SINR}_{m}>\tau|\Phi_m\neq\emptyset,\Phi_s\neq\emptyset,r_{\min,s}>\rho{}r_{\min,m}\right)\notag\\
&\times\Pr\left(\Phi_m\neq\emptyset,\Phi_s\neq\emptyset,r_{\min,s}>\rho{}r_{\min,m}\right).
\end{align}
Following the derivation procedure of $P_2\left(\tau\right)$, we obtain $P_4\left(\tau\right)$ as
\begin{align}\label{P4}
P_4\left(\tau\right)
=&B_m B_s \int_{0}^{\mu_m}\left(1-F_{r_{\min,s}}\left(\rho x\right)\right)\sum_{k=0}^{\mathcal{M}-1}\frac{\left(-\psi_{m}x^{\alpha}\right)^k}{k!}\notag\\
&\times\mathcal{L}_{I_m^{\ast}+I_s+\sigma^{2}}^{\left(k\right)}\left(\psi_{m}x^{\alpha}\right) f_{r_{\min,m}}\left(x\right)dx.
\end{align}
Substituting \eqref{Frmins} and \eqref{frmins} in \eqref{P4}, we further derive $P_4\left(\tau\right)$ as
\begin{align}\label{P4L}
P_4\left(\tau\right)
=&B_{m}B_{s}\int_{0}^{\min\left(\frac{\mu_s}{\rho},\mu_m\right)}\left(1-\frac{1-e^{-\lambda_s \pi x^2\rho^2}}{B_s}\right)\notag\\
&\times\sum_{k=0}^{\mathcal{M}-1}\frac{\left(-\psi_{m}x^{\alpha}\right)^k}{k!}
\mathcal{L}_{I_m^{\ast}+I_s+\sigma^{2}}^{\left(k\right)}\left(\psi_{m}x^{\alpha}\right)\notag\\
&\times\frac{2\lambda_{m}\pi{}x{}e^{-\lambda_{m}\pi{}x^2}}{B_m}dx,
\end{align}
where
\begin{equation}
\mathcal{L}_{I_m^{\ast}+I_s+\sigma^{2}}\left(a\right)
=e^{-a\sigma^{2}}e^{-\lambda_m\int_{x}^{\mu_m}r\Omega_{m}dr
-\lambda_s\int_{0}^{\mu_s}r\Omega_{s}dr}.
\end{equation}

\subsubsection{Analysis of $P_5\left(\tau\right)$}

In Scenario 5, the power of the received signal at the typical UE is given by $S=\kappa_{s}\hbar$
and power of the interference at the typical UE is given by $I=I_m+I_s^{\ast}$, where $I_s^{\ast}$ is defined in \eqref{Isast} and $I_{m}$ is given by
\begin{equation}\label{Im}
I_{m}\triangleq\sum_{\left(r_m,\theta\right)\in\Phi_m}
P_{m}r_{m}^{-\alpha}G_m\left(\theta\right)\hbar.
\end{equation}
Following the derivation procedure of $P_4\left(\tau\right)$, we derive $P_5\left(\tau\right)$ as
\begin{align}\label{P5L}
P_5\left(\tau\right)=&B_{m}B_{s}\int_{0}^{\min\left(\mu_s,\rho\mu_m\right)} \left(1-\frac{1-e^{-\frac{\lambda_{m}\pi{}x^{2}}{\rho^{2}}}}{B_m}\right)\notag\\
&\times\sum_{k=0}^{\mathcal{M}-1}\frac{\left(-\psi_{s}x^{\alpha}\right)^k}{k!}
\mathcal{L}_{I_m+I_s^{\ast}+\sigma^{2}}^{\left(k\right)}\left(\psi_{s}x^{\alpha}\right)\notag\\
&\times\frac{2\lambda_s\pi{}x{}e^{-\lambda_{s}\pi{}x^2}}{B_s}dx,
\end{align}
where
\begin{equation}
\mathcal{L}_{I_m+I_s^{\ast}+\sigma^{2}}\left(a\right)
=e^{-a\sigma^{2}}e^{-\lambda_m\int_{0}^{\mu_m}r\Omega_{m}dr
-\lambda_s\int_{x}^{\mu_s}r\Omega_{s}dr}.
\end{equation}

\begin{table}[!t]
\centering
\caption{Parameters Used in Section~\ref{sec:numerical}}\label{table:SimulationParameter}
\vspace{-2pt}
\begin{tabular}{|l|l|}
\hline
Parameters                                  & Value                          \\ \hline\hline
$P_m$ \& $P_s$                              & $10^4$ \& $10^2$ mW               \\ \hline
$\lambda_m$, $\lambda_s$, \& $\lambda_u$    & $10^{-5}$, $10^{-4}$, \& $10^{-1}$/m$^2$ \\ \hline
$\mu_m$ \& $\mu_s$                          & $1000$ \& $100$m                \\ \hline
$\omega_m$ \& $\omega_s$                    & $0.6$ \& $0.5$                   \\ \hline
$\theta_m$ \& $\theta_s$                    & $0.1$ \& $0.2$ rad               \\ \hline
$G_{\max,m}$ \& $G_{\max,s}$                & $4\times10^3$ \& $10^3$               \\ \hline
$G_{\min,m}$ \& $G_{\min,s}$                & $1$ \& $1$                     \\ \hline
$A_s$                                       & $100$                          \\ \hline
$\alpha$                                    & $2.2$                          \\ \hline
$\mathcal{M}$                                         & $1$                            \\ \hline
$\sigma^2$                                  & $1$                            \\ \hline
$S_m$ \& $S_s$                              & $10^9$ \& $10^9$ Hz            \\ \hline
\end{tabular}
\end{table}

Substituting the rate coverage probabilities derived in \eqref{P2L}, \eqref{P3L}, \eqref{P4L}, and \eqref{P5L} into \eqref{Rate}, we obtain the rate coverage probability of the network, $P_{c}$.
The expression for $P_{c}$ reveals the impact of the bias factor, $A_s$, on the rate coverage probability of the network. We note that a closed-form expression for the optimal $A_s$ which achieves the highest rate coverage probability of the network is mathematically intractable. To address this, we can find the optimal $A_s$ via the linear search method, as will be shown in Fig. \ref{fig:numerical:6}.

\section{Numerical Results and Discussions}\label{sec:numerical}

In this section, we present numerical results to examine the impact of the mmWave cellular HetNet parameters on the network performance. Based on these results, we provide some useful insights into the design of the mmWave cellular HetNet for achieving the best network performance. Unless otherwise specified, the parameters used in this section are summarized in Table~\ref{table:SimulationParameter}.

\begin{figure}[!t]
    \begin{center}
        \includegraphics[height=2.2in,width=3.1in]{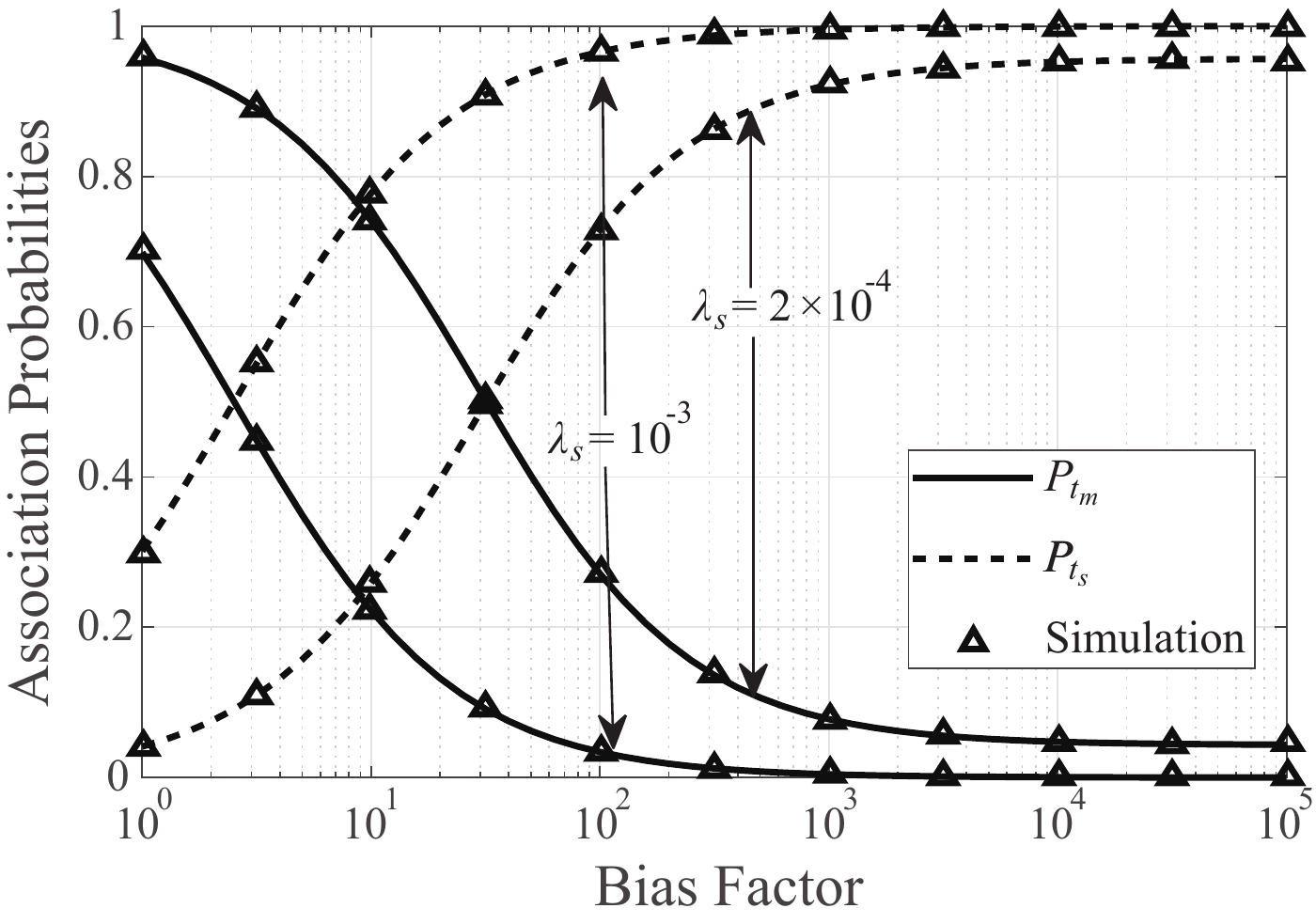}
        \caption{The association probabilities, $P_{t_{m}}$ and $P_{t_{s}}$, versus the bias factor, $A_{s}$, for different values of $\lambda_{s}$.}\label{fig:numerical:1}
    \end{center}
\end{figure}

We first evaluate the impact of the LoS low-power BS intensity, $\lambda_{s}$, and the bias factor, $A_s$, on the association probabilities, $P_{t_{m}}$ and $P_{t_{s}}$. In Fig.~\ref{fig:numerical:1}, we plot $P_{t_{m}}$ and $P_{t_{s}}$ versus $A_{s}$ for $\lambda_{s}=2\times10^{-4}/m^2$ and $\lambda_{s}=10^{-3}/m^2$. By comparing the analytical curves with the Monte Carlo simulation points marked by `$\vartriangle$', we observe that our analytical expressions for $P_{t_{m}}$ and $P_{t_{s}}$, given in \eqref{PtM} and \eqref{PtS}, precisely agree with the simulations, which corroborates the accuracy of our analysis. Moreover, we observe that when $A_s$ increases, $P_{t_{s}}$ increases while $P_{t_{m}}$ decreases. This observation is not surprising since increasing $A_s$ leads to a lower probability that \eqref{condition_macro_choose} holds, i.e., a lower probability that the typical UE chooses a macro BS as its serving BS. Furthermore, we observe that the limit value of $P_{t_{m}}$ as $A_{s}$ grows large for $\lambda_{s}=10^{-3}/m^2$ is lower than that for $\lambda_{s}=2\times10^{-4}/m^2$. This can be analytically explained based on \eqref{PtM}. Specifically, we find from \eqref{PtM} that $\lim_{A_{s}\to\infty}P_{t_{m}}=B_{m}\left(1-B_{s}\right)
=\Pr\left(\Phi_m\neq\emptyset\right)\Pr\left(\Phi_s=\emptyset\right)$. When $\lambda_{s}$ increases, there are more LoS low-power BSs available in the network and thus, $P_{t_{m}}$ decreases. The observation that the limit value of $P_{t_{s}}$ as $A_{s}$ grows large for $\lambda_{s}=10^{-3}/m^2$ is higher than that for $\lambda_{s}=2\times10^{-4}/m^2$ can be explained in a similar fashion. Finally, we observe that $P_{t_{m}}+P_{t_{s}}=1-\Pr\left(\textrm{Scenario~1}\right)$ for a given bias factor, which is under expectation. Of course, we note that $\Pr\left(\textrm{Scenario~1}\right)$ is very small in Fig.~\ref{fig:numerical:1}.

\begin{figure}[!t]
    \begin{center}
        \includegraphics[height=2.2in,width=3.1in]{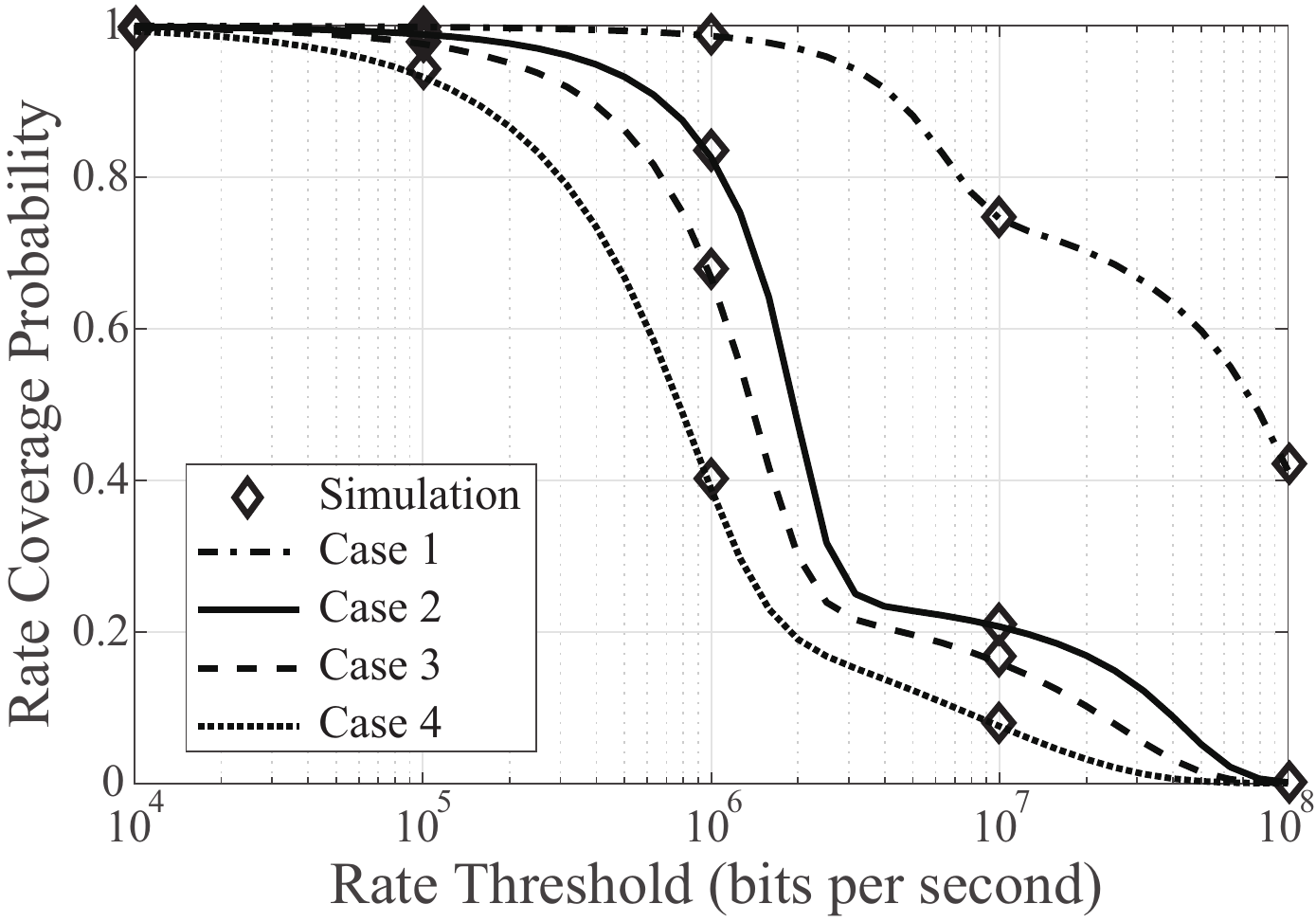}
        \caption{The rate coverage probability, $P_c$, versus the rate threshold, $\delta$, for four cases: Case 1: $\lambda_s=10^{-3}/m^2$, $\theta_m=0.1$ rad, and $\theta_s=0.2$ rad, Case 2: $\lambda_s=10^{-4}/m^2$, $\theta_m=0.1$ rad, and $\theta_s=0.2$ rad, Case 3: $\lambda_s=10^{-4}/m^2$, $\theta_m=0.2$ rad, and $\theta_s=0.4$ rad, and Case 4: $\lambda_s=10^{-4}/m^2$, $\theta_m=0.5$ rad, and $\theta_s=1$ rad.}
        \label{fig:numerical:4}
    \end{center}
\end{figure}

Second, we evaluate the impact of $\lambda_s$ and the beamwidth of the BS, $\theta_m$ and $\theta_s$, on the rate coverage probability, $P_{c}$. In Fig. \ref{fig:numerical:4}, we plot $P_{c}$ versus the rate threshold, $\delta$, for four cases with different values of $\theta_m$, $\theta_s$, and $\lambda_s$. First, we demonstrate the correctness of our analytical expression for $P_{c}$ by the exact match between the analytical curves and the Monte Carlo simulation points marked by `$\diamond$'. Second, by comparing Case 2 with Cases 3 and 4, we find that $P_{c}$ improves when the beam becomes narrower. This observation is expected, since the narrower the beam, the less interference caused by BSs. Third, by comparing Case 1 with Case 2, we find that deploying more low-power BSs significantly improves the rate coverage performance when beams are narrow.

\begin{figure}[!t]
    \begin{center}
        \includegraphics[height=2.2in,width=3.1in]{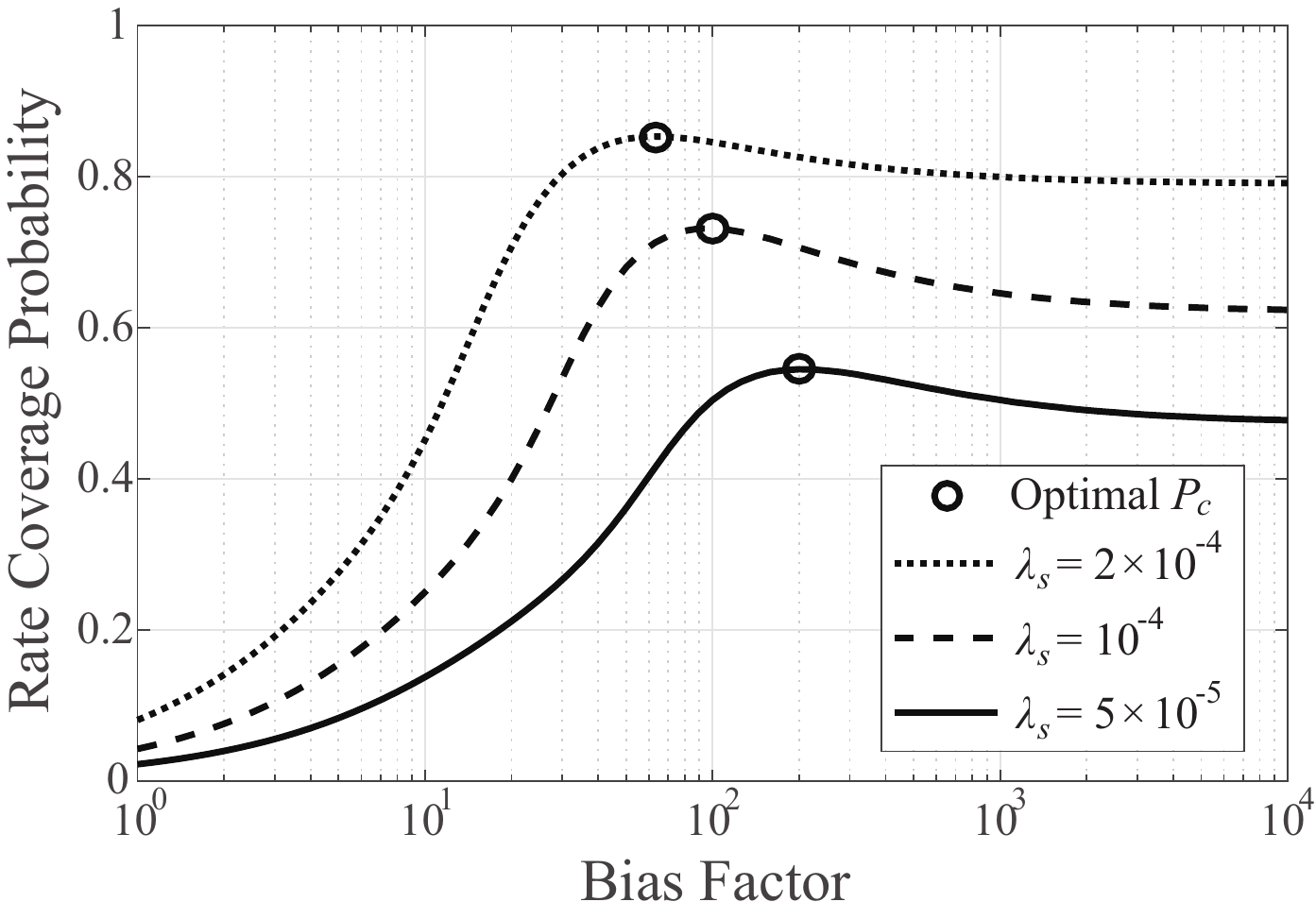}
        \caption{The rate coverage probability, $P_c$, versus the bias factor, $A_s$, for different values of $\lambda_s$ with $\delta=10^{6.5}$ bits per second.}
        \label{fig:numerical:6}
    \end{center}
\end{figure}

Third, we evaluate the impact of $\lambda_{s}$ and $A_{s}$ on $P_{c}$. In Fig. \ref{fig:numerical:6}, we plot $P_{c}$ versus $A_s$ for different $\lambda_s$. First, we find that given $\lambda_{s}$, $P_{c}$ first increases, then decreases, and finally becomes saturated when $A_{s}$ increases. This demonstrates the existence of the optimal value of $A_{s}$ which maximizes $P_{c}$ for a given $\lambda_{s}$. This confirms our statement in the last paragraph of Section~\ref{sec:coverage_analysis} that the optimal $A_{s}$ maximizing $P_{c}$ can be found via the linear search method. Second, we compare the value of $P_{c}$ when $A_{s}=1$ with the value of $P_{c}$ when $A_{s}>1$ and find that the use of the CRE technique, which offloads more UEs to low-power BSs, significantly improves the rate coverage performance. Third, we observe that the optimal $A_s$ increases when $\lambda_{s}$ decreases. This observation is expected since when the intensity of low-power BSs becomes lower, more UEs tend to choose the macro-BSs. In this case, a larger $A_s$ is needed to offload UEs from macro BSs to low-power BSs.

\begin{figure}[!t]
    \begin{center}
        \includegraphics[height=2.2in,width=3.1in]{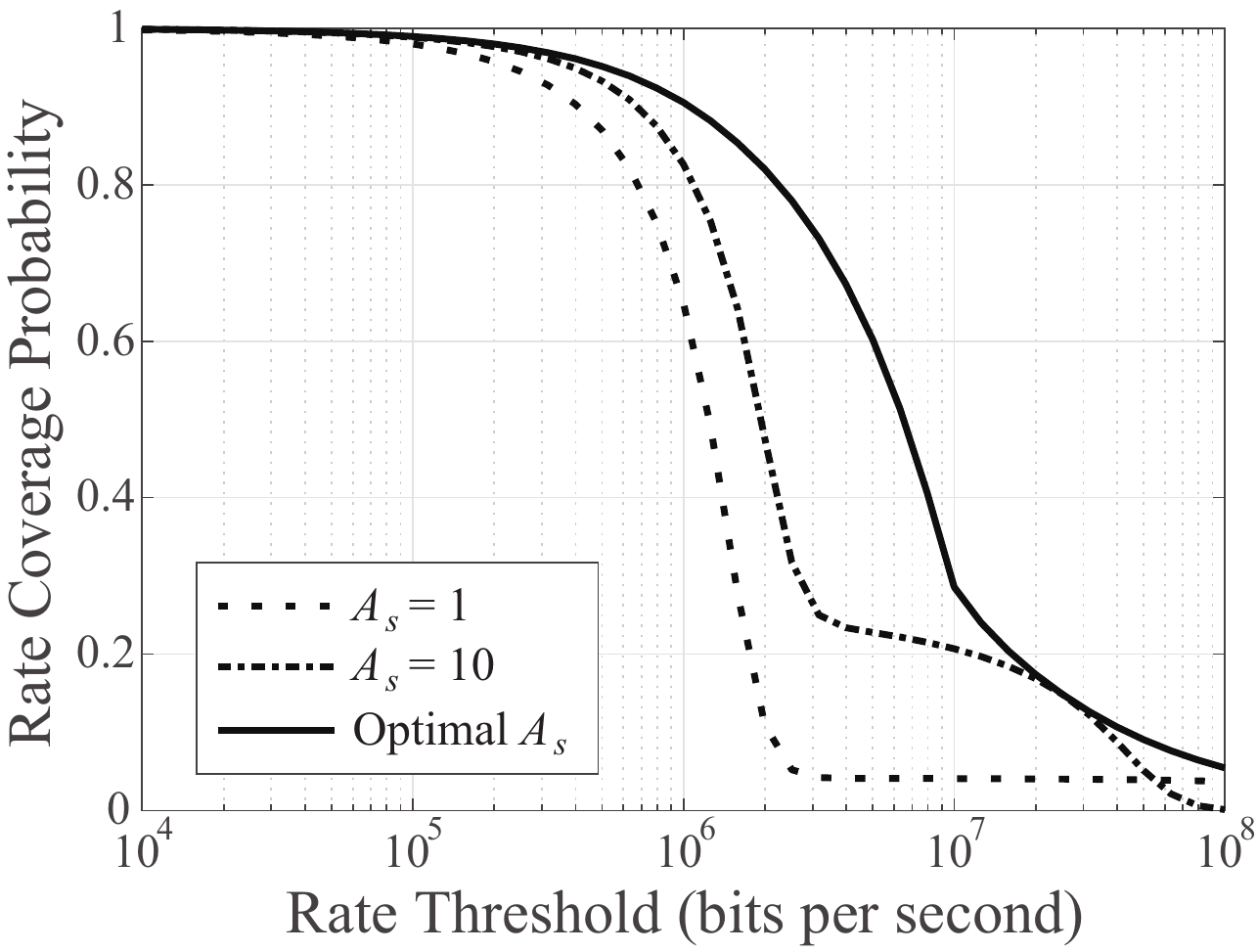}
        \caption{The rate coverage probability, $P_c$, versus the rate threshold, $\delta$, for different values of $A_s$.}
        \label{fig:numerical:7}
    \end{center}
\end{figure}

Finally, we investigate the effectiveness of the optimal $A_s$. In Fig. \ref{fig:numerical:7}, we compare the optimal rate coverage probability achieved by using the optimal $A_s$ with those achieved by setting $A_s=1$ and $A_s=10$ versus $\delta$. Clearly, we find that the optimal rate coverage probability is always higher than those with $A_s=1$ and $A_s=10$, across the whole range of $\delta$. In particular, the rate coverage performance advantage achieved by the optimal $A_s$ over the fixed $\delta$ is more profound when $\delta$ is in the medium regime, e.g., $10^{6}\leq\delta\leq10^{7}$. This demonstrates that the optimal $A_s$ found through our analysis effectively and significantly improves the rate coverage performance of the mmWave cellular HetNet.

\section{Conclusions}\label{sec:conclusion}

In this contribution, we proposed an effective method to optimize the bias factor in a two-tier mmWave cellular HetNet. Importantly, the unique features of mmWave networks, e.g., extremely narrow beams of BSs and vulnerability to blocking, were addressed by adopting the sectored antenna gain model and the LoS ball model, respectively. For this mmWave HetNet, we first analyzed the loads of the macro BS and the low-power BS. Then we derived a new expression for the rate coverage probability experienced by the typical UE in the network. Based on this expression, we determined the value of the optimal bias factor which maximizes the rate coverage probability through linear search. With numerical results, we showed that the determined optimal bias factor can significantly improve the rate coverage probability compared to a fixed bias factor. In addition, the impact of various network parameters, e.g., the densities and the beamwidths of BSs, on the rate coverage probability was examined.

\end{document}